# Generative Adversarial Network for Personalized Art Therapy in Melanoma Disease Management

**Lennart Jütte,[a,*] Ning Wang,[a] Bernhard Roth[a,b]**
[a]Leibniz University Hannover, Hannover Centre for Optical Technologies, Hannover, Germany
[b]Leibniz University Hannover, Cluster of Excellence PhoenixD, Hannover, Germany

**Abstract**

**Significance:** Melanoma is the most lethal type of skin cancer. Patients are vulnerable to mental health illnesses which can reduce the effectiveness of the cancer treatment and the patients' adherence to drug plans. It is crucial to preserve the mental health of patients while they are receiving treatment. However, current art therapy approaches are not personal and unique to the patient.

**Aim:** We aim to provide a well-trained image style transfer model that can quickly generate unique art from personal dermoscopic melanoma images as an additional tool for art therapy in disease management of melanoma.

**Approach:** Visual art appreciation as a common form of art therapy in disease management that measurably reduces the degree of psychological distress. We developed a network based on the cycle-consistent generative adversarial network for style transfer that generates personalized and unique artworks from dermoscopic melanoma images.

**Results:** We developed a model that converts melanoma images into unique flower-themed artworks that relate to the shape of the lesion and are therefore personal to the patient. Further, we altered the initial framework and made comparisons and evaluations of the results. With this, we increased the options in the toolbox for art therapy in disease management of melanoma. The development of an easy-to-use user interface ensures the availability of the approach to stakeholders.

**Conclusions:** The transformation of melanoma into flower-themed artworks is achieved by the proposed model and the graphical user interface. This contribution opens a new field of GANs in art therapy and could lead to more personalized disease management.

**Keywords**: melanoma, generative adversarial network, image style transfer, disease management, mental health.

*Lennart Jütte, E-mail: lennart.juette@hot.uni-hannover.de

## 1 Introduction

Melanoma accounts for the most deaths from any skin cancer[1] and develops from the pigment-producing cells known as melanocytes[2]. The incidence has been increasing over the past 30 years[4]. While suffering from physical pain, patients are also vulnerable to mental illnesses. Guy GP Jr et al. show that about 30% of all patients diagnosed with melanoma report on psychological issues[5]. This is a problem because follow-up research indicates that mental health problems can reduce the effectiveness of cancer medication[6,7]. Therefore, it is essential to care for the mental health of the patient during treatment as well. Even after successful treatment, melanoma survivors face psychological problems. For example, they may feel anxious or depressed due to fear of living with the disease or cancer recurrence[8]. In conclusion, adequate psychological help is necessary to achieve better coping with the illness.

Art therapy is an approach for such psychological intervention that focuses on the mental health aspects of the patient and reduces the degree of psychological distress. It is used to offer support to cancer patients at different stages of their illness. It has been shown to improve the patient's self-consciousness and resilience to pressure[9,10,11]. Visual art appreciation of famous paintings, such as *Starry Night* or *Sunflowers* by Vincent Van Gogh, is a common form of art therapy, and cancer patients feel relaxed in their emotional state through visual art appreciation[12,13].

With the development of deep learning, Goodfellow et al., in June 2014, designed a machine learning framework, the Generative Adversarial Network (GAN)[14]. Subsequently, more researchers started to propose a series of variants of GANs for different problems in various fields. Regarding the image style transfer, Isola et al. proposed the pix2pix algorithm model[15], which can learn the mapping relationship between images of different domains without supervision. However, it can only use paired datasets for training. To solve this problem, Zhu et al. proposed the CycleGAN model utilizing cycle consistency loss to constrain training and achieve cross-domain image transformation with unpaired datasets[16].

In this work, we propose the transformation of melanoma images into art paintings based on an adjusted variant of the CycleGAN model. With this work we aim to provide a novel tool for art therapy that enables easier access to art therapy for patients because the generated artworks are personal and unique to each disease. GANs are ideal for this concept since, following the initial network training, image production is automated. This contribution to the field of personalized medicine and personalized therapy further expands the options and tools for art therapists in the disease management of melanoma and could help to avoid a reduction of the effectiveness of the treatment.

## 2　Methods

*2.1 Basic Cycle-Consistent GAN Working Principle*

The style transfer model used in this work is based on the cycle-consistent GAN framework proposed by Zhu et al[16]. The structure consists of two GANs, as shown in Figure 1.

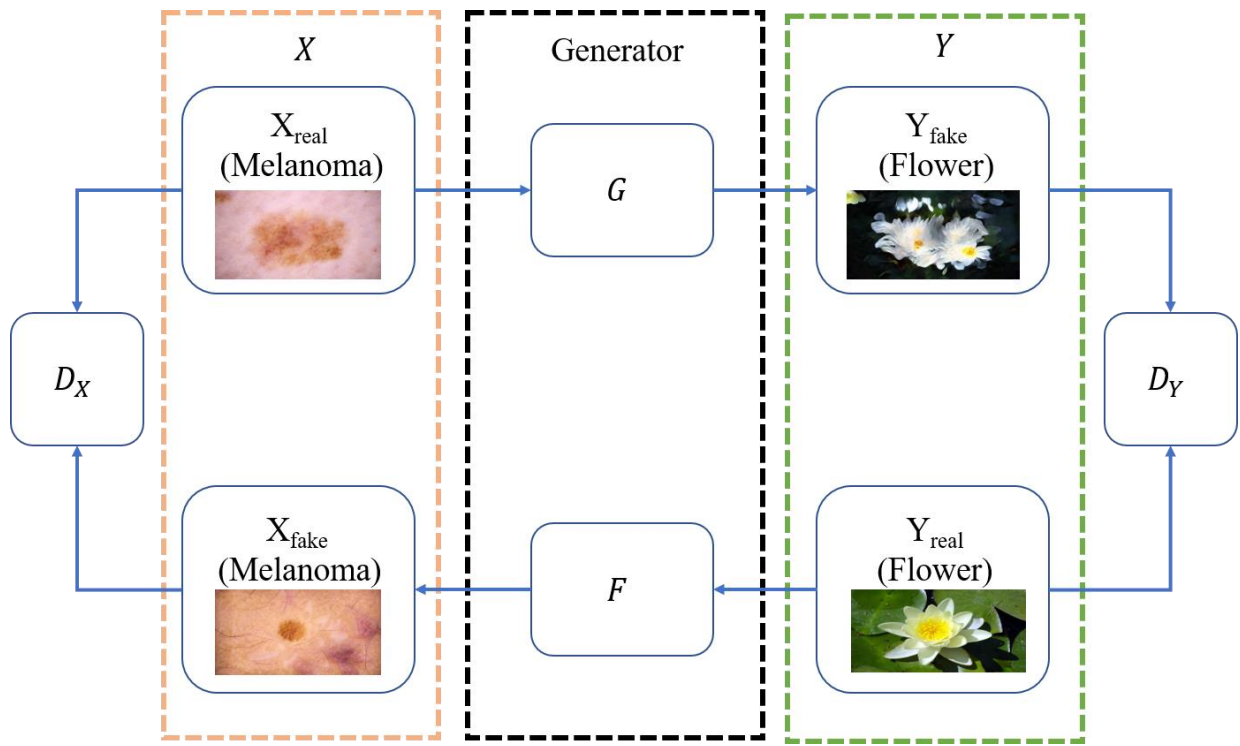

**Fig. 1** Cycle consistent GAN model. It contains two generators, $\boldsymbol{G}$ and $\boldsymbol{F}$, and two discriminators, $\boldsymbol{D_X}$ and $\boldsymbol{D_Y}$. Melanoma image as the source domain $\boldsymbol{X}$, flower image as the target domain $\boldsymbol{Y}$.

$X$ and $Y$ represent images from two different domains. In this work, the source domain $X$ is the melanoma image, and the target domain $Y$ is the flower image. $G$ and $F$ are the two generators. Here, $G$ converts the melanoma image to the flower image, and $F$ is the opposite conversion. In addition, there are two discriminators, $D_X$ and $D_Y$, employed to determine whether an image belongs to a specific domain. In this work, $D_Y$ will be utilized to judge whether an image is an image of the flower, whether it is a fake flower image generated from $G$ or a real flower image that is already in the flower dataset $Y$. The real flower images as well as the synthetically generated flower images will be judged by $D_Y$.

*2.2 Generator Architecture*

The two generators, $G$ and $F$, have the same network structure in this work. The generator network contains downsampling, residual blocks, and upsampling. The framework of the generator is shown in Figure 2.

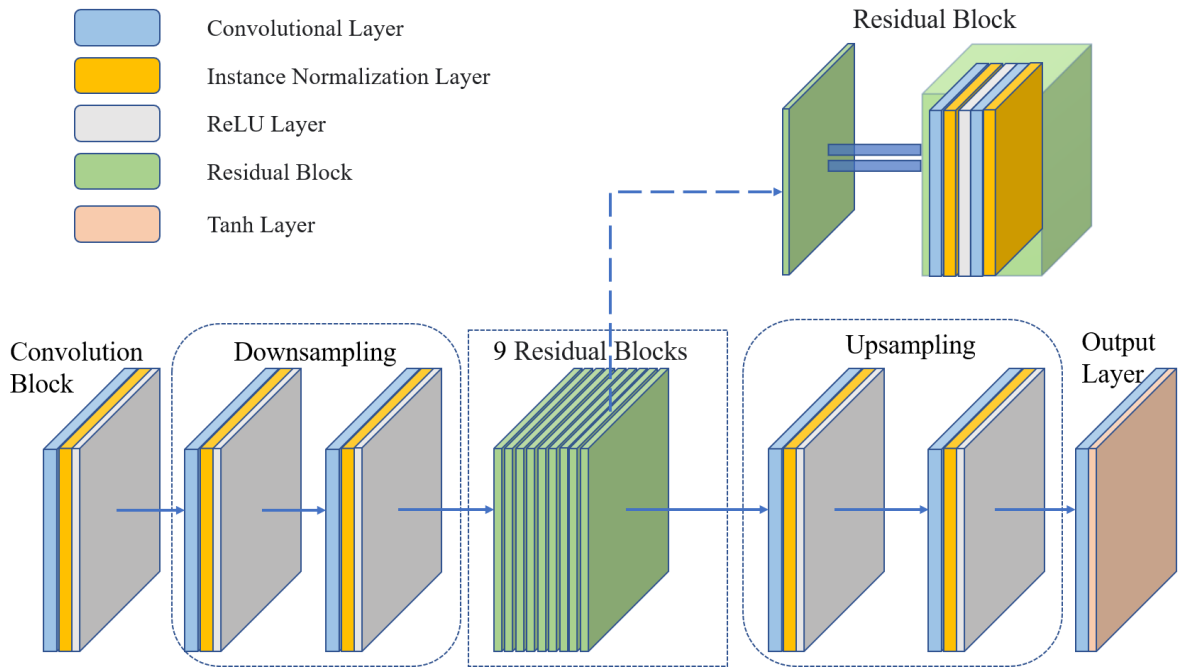

**Fig. 2** Structural diagram of the generator. The generator network consists of a convolution block, a downsampling block, nine residual blocks, an upsampling block and an output layer.

In this work, the input image size of the generator is 3*256*256. The first layer is a convolutional layer with 64 output channels. The input will be padded to 262*262, i.e., padding size of 3, and convoluted by a kernel of size 7*7 with the stride of 1, resulting in 64 features of size 256*256. Afterwards it is then processed using the Instance Normalization (IN) and ReLU activation functions. The following two convolution layers downsample the image, reduce the image size and increase the number of channels. Like cycleGAN, many residual blocks are built in the generator network to avoid degradation in deep networks[17]. In this work, we use nine residual blocks with the number of channels remaining the same. Subsequently, upsampling is performed to reduce the number of channels and restore to the original image size. The last convolution layer utilizes the hyperbolic tangent as the activation function. Table 1 shows the parameters and the size of the output for each convolution layer of the generator. $N$ is the number of channels in the convolutional layer, $K$ is the size of the convolution kernel, $S$ is the stride, and $P$ is the padding size.

**Table 1** The parameters and the size of the output for each convolutional layer of the generator. *N* is the number of channels in the convolutional layer, *K* is the size of the convolution kernel, *S* is the stride, and *P* is the padding size. The input size in this work is 3*256*256(*C* Channel * *H* Height * *W* Width).

| Component | Convolution Layer Parameter | | | | | Output Size | | |
|---|---|---|---|---|---|---|---|---|
| | Type | N | K | S | P | C | H | W |
| Convolution Block | Conv | 64 | 7 | 1 | 3 | 64 | 256 | 256 |
| Downsampling | Conv | 128 | 3 | 2 | 1 | 128 | 128 | 128 |
| | Conv | 256 | 3 | 2 | 1 | 256 | 64 | 64 |
| 9 Residual Blocks | Res | 256 | 3 | 2 | 1 | 256 | 64 | 64 |
| Upsampling | Conv | 128 | 3 | 2 | 1 | 128 | 128 | 128 |
| | Conv | 256 | 3 | 2 | 1 | 64 | 256 | 256 |
| Output Layer | Conv | 3 | 7 | 1 | 3 | 3 | 256 | 256 |

*2.3 Discriminator Architecture*

Different from other discriminators, which map the input to a real number as a representation of the probability of belonging to the real image, the patchGAN maps the input to a matrix of size N*N, i.e., the patch[15]. Each value in the matrix represents the discriminative result of a small receptive field of the image. The discriminator network in this work uses a patchGAN with a receptive field of 70*70. The discriminator structure is shown in Figure 3.

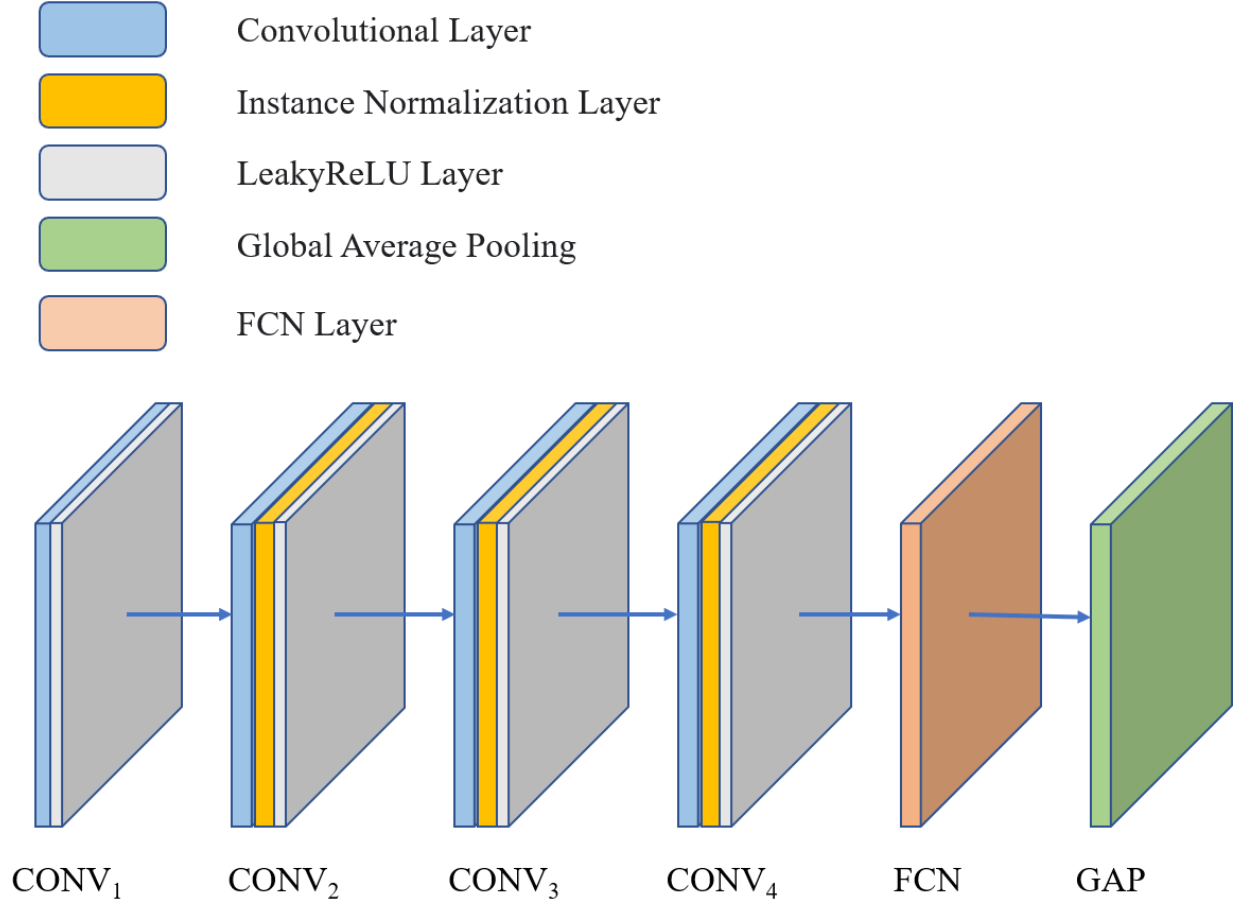

**Fig. 3** Structural diagram of the discriminator. The discriminator network consists of four convolutional blocks (CONV) that perform convolutional operations to extract image features, a fully connected layer (FCN) to combine the features extracted from the previous convolutional layers, and a global average pooing layer (GAP) to reduce the number of the parameters.

The kernel size of the first three convolutional layers is 4*4 with stride 2 and padding 1, which convert the input from 3*256*256 to 256*32*32. The kernel size of the following convolutional layer and the fully connected layer is 4*4 with stride 1 and padding 1, converting the input to 512*31*31 and 1*30*30. Finally, the global average pooling layer reduces the large number of parameters[18], making the model more robust and resistant to overfitting. The discriminator activation function is leaky ReLU[19], which keeps a minimal gradient when the input is negative to avoid gradient disappearance. Table 2 shows the parameters and the size of the output for each convolutional layer of the discriminator.

**Table 2** The parameters and the size of the output for each convolutional layer of the discriminator, the input size is 3*256*256(Channel*Height*Width). *N* is the number of channels in the convolutional layer, *K* is the size of the convolution kernel, *S* is the stride, and *P* is the padding size. The output size is given as per *C* Channel * *H* Height * *W* Width.

| Component | Convolutional Layer Parameter | | | | | Output Size | | |
|---|---|---|---|---|---|---|---|---|
| | Type | N | K | S | P | C | H | W |
| **CONV₁** | Conv | 64 | 4 | 2 | 1 | 64 | 128 | 128 |
| **CONV₂** | Conv | 128 | 4 | 2 | 1 | 128 | 64 | 64 |
| **CONV₃** | Conv | 256 | 4 | 2 | 1 | 256 | 32 | 32 |
| **CONV₄** | Conv | 512 | 4 | 2 | 1 | 512 | 31 | 31 |
| **FCN** | Conv | 1 | 4 | 1 | 1 | 1 | 30 | 30 |

*2.4 Network Structure Adjustments*

We adjusted the above network structure to the demands in the application of this work. In the following, we explain the adjusted structure of the network with the integration of a sub-pixel convolution, an attention mechanism, and a spectral normalization.

*2.4.1 Sub-Pixel Convolution*

We employ sub-pixel convolution instead of transposed convolution for the upsampling in the generator. The transposed convolution is implemented by padding the image with zero then performing convolution, while sub-pixel convolution generates multiple channels by convolution and then reshapes them[20]. The advantage of the sub-pixel convolution is that it has a larger receptive field, provides more contextual information, and can generate more details[21]. This is useful for obtaining high-resolution results. The generator architecture with the sub-pixel convolution layer is shown in Figure 4. The sub-pixel convolution block consists of a convolution layer with 1024 output features, a pixel shuffle layer with a factor of 4, an instance normalization layer, and a ReLU activation function layer.

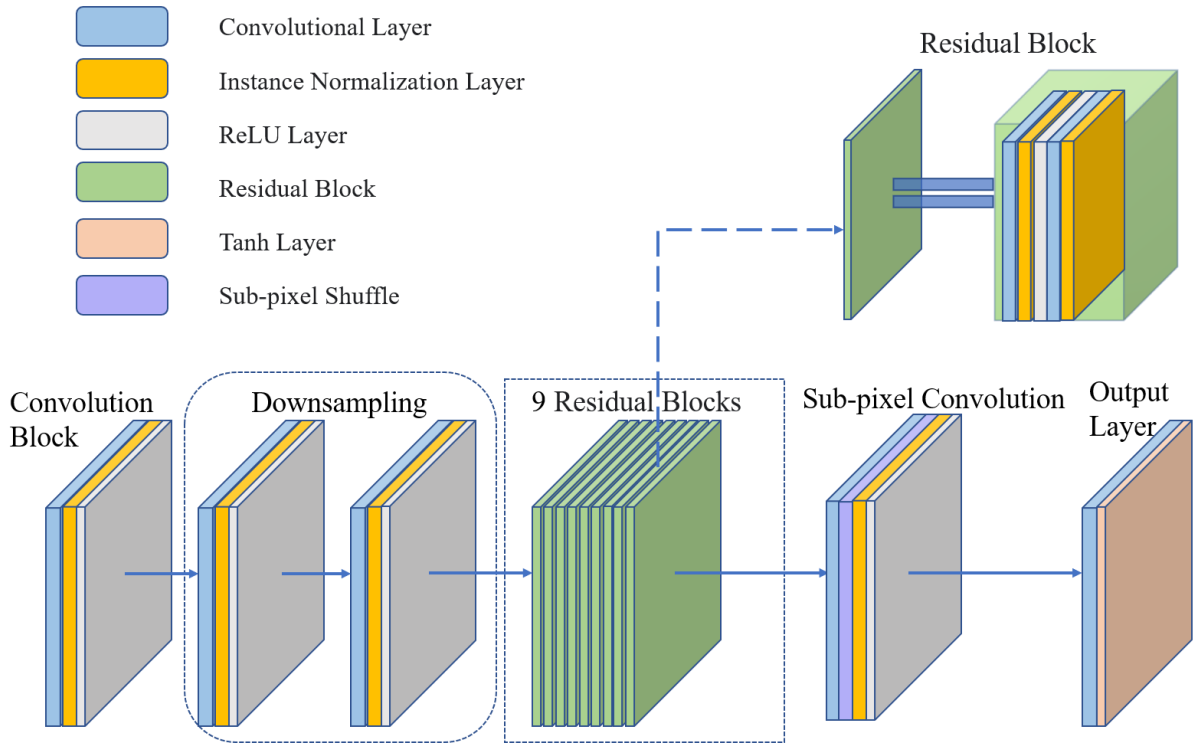

**Fig. 4** Structural diagram of the generator with sub-pixel convolution. Extracting features with a convolutional block and downsampling, residual blocks help to avoid the neural network degradation. We employ sub-pixel convolution to complete the upsampling.

*2.4.2 Attention Mechanism*

In deep learning, the attention mechanism[22] is utilized to obtain distant dependencies. In the image task, the distant dependencies are the receptive fields formed by convolutional operations. Wang et al. proposed a self-attention mechanism for the problem that traditional convolutional operations can only process local information to coordinate the details of the image relatively globally[23]. The structure of the self-attention mechanism is shown in Figure 5.

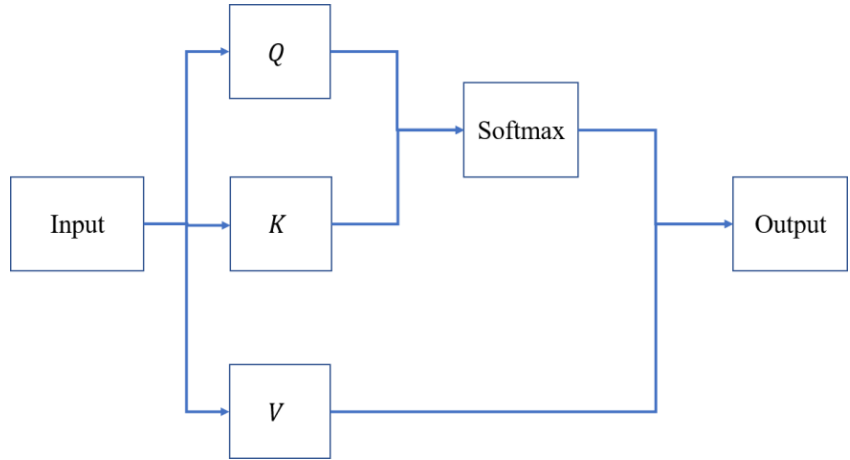

**Fig. 5** Structure of the self-attention mechanism. The three vectors are query (***Q***), key (***K***) and value (***V***). ***Q*** is the query vector related to the encoded information, while ***K*** is the output vector of the encoder. ***V*** is the vector related to the input. With ***Q*** and ***K*** the attention can be calculated to perform the Softmax and then multiply it with ***V***. The input here is the dermoscopic image.

$Q$ (query), $K$ (key) and $V$ (value) are the features obtained by the linear mapping of the input feature map. First, the matrices of $Q$ and $K$ are multiplied to obtain the attention matrix. Afterwards, the Softmax operation is performed and then the resulting attention is applied to $V$ to obtain the attention feature[22]. In this work, we add one self-attention layer for the generator after the downsampling part as well as after the residual blocks. Figure 6 shows the network structure of the generator including the attention mechanism.

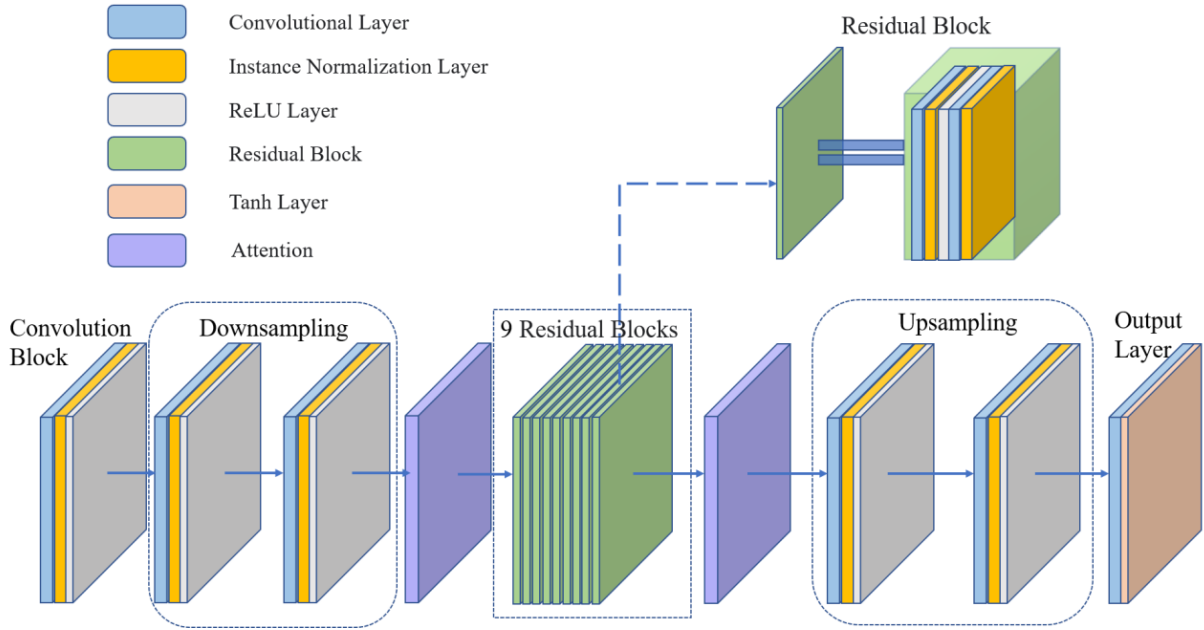

**Fig. 6** Structural diagram of the generator with self-attention. There is one self-attention layer after the downsampling and the residual blocks, respectively.

*2.4.3 Spectral Normalization*

Spectral normalization was proposed by Miyato et al. and applied to discriminator networks, it stabilizes the training. It is challenging to train a GAN because occasionally the discriminator reaches the optimal state extremely quick, making it unable to train the generator more effectively. We can solve this problem with spectral normalization. It can limit the Lipschitz constant of the discriminator to improve control of the discriminative network while limiting the upper gradient to make it less prone to gradient explosion[24]. The structure of the discriminator network with the spectral normalization and the self-attention is illustrated in Figure 7.

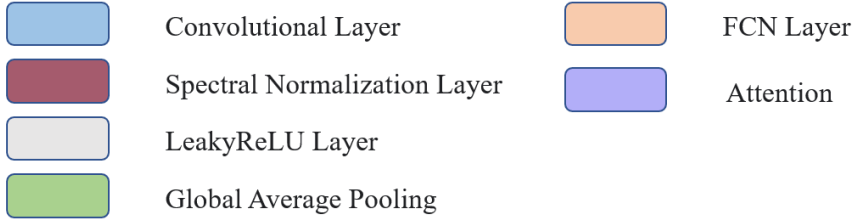

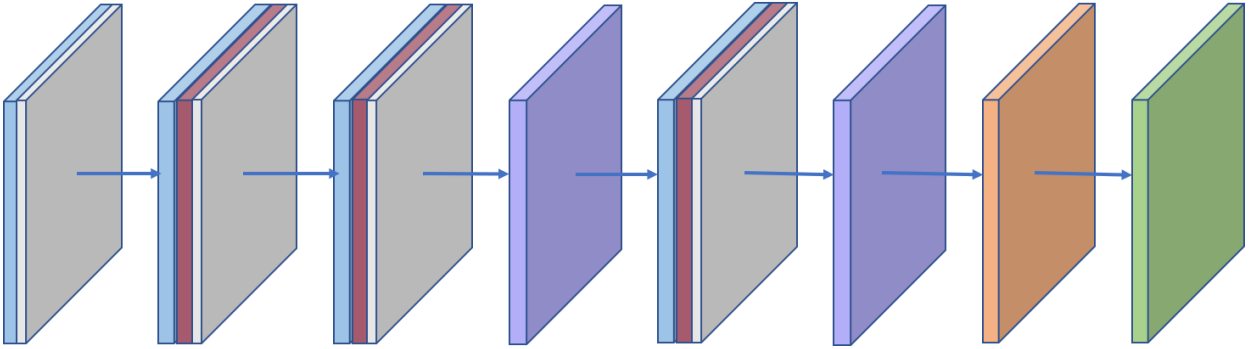

**Fig. 7** Structural diagram of discriminator with self-attention and spectral normalization. We replaced the instance normalization with spectral normalization and add self-attention after the last two convolution blocks compare to the discriminator introduced in chapter 2.3.

## 2.5 Loss Function

We employ three loss functions in this work: adversarial loss, cycle-consistency loss, and identity loss. In the following, these loss functions are explained in detail.

### 2.5.1 Adversarial Loss

The purpose of using adversarial loss is to ensure the image generated by the generator resembles the distribution of real data[14]. As opposed to the original CycleGAN proposed by Zhu et al. which used cross-entropy[16], we apply the least squares loss function which has better robustness for outliers as the adversarial loss function[25]. The equation is as:

$$L^{s \to t} = E_{x \sim X_t}[(D_t(X))^2] + E_{x \sim X_s}[(1 - D_t(G(X)))^2] \qquad (1)$$

$S$ is the source domain and $T$ the target domain. $x \sim X_t$ represents the sample from the target domain and $x \sim X_s$ stands for the sample from the source domain. $E$ means expectation. $D$ is the discriminator. The output value is 0 or 1, representing whether the input $X$ is from the generated data or the real data. The generator $G$ converts the images from the source domain into the target domain.

*2.5.2 Cycle-Consistency Loss*

Cycle consistency ensures that the results generated by the two generators do not contradict each other. A melanoma image $X$ is converted to a flower image $Y_f$ by generator $G$. After that, $Y_f$ is converted back to a melanoma image $X_r$ by generator $F$. The contents in the two images of $X$ and $X_r$ should ideally be the same. The cycle consistency loss function is defined by the $L^1$-norm of these two images, $X$ and $X_r$. The equation is as:

$$L_{cycle}^{s \to t} = E_{x \sim X_s}[\|X - F(G(X))\|_1] \qquad (2)$$

$S$ is the source domain. $T$ is the target domain. $x \sim X_s$ represents the sample from the source domain and $E$ stands for expectation. $G$ is the generator which converts the input $X$ from source domain into target domain. In this work, $G$ converts melanoma images into flower images, $F$ is another generator which converts the flower images into melanoma images.

*2.5.3 Identity Loss*

Generator $G$ converts melanoma images to flower images. If we use a flower image $Y$ as input to the generator $G$, the output $Y_f$ should not change. This is the purpose of using identity loss. We calculate the $L1$ norm of $Y$ and $Y_f$ with the following equation:

$$L_{ide}^{s \to t} = E_{x \sim X_t}[\|Y - G(Y)\|_1] \qquad (3)$$

$S$ is the source domain. $T$ is the target domain. $x \sim X_t$ represents the sample from the target domain. $E$ means expectation. $G$ is the generator which converts the input from source domain into target domain, i.e., melanoma images into flower images.

*2.6 Datasets and Training Parameters*

In this work, we use two public datasets. The melanoma image dataset is the official dataset of the SIIM-ISIC Melanoma Classification Challenge 2020[26], and the flower image dataset is the Oxford 102 Flower dataset[27]. Considering the hardware limitations in this work, we randomly selected 3661 melanoma images and 2079 images of flowers. For training, the batch size is 1, and the number of iterations is 200, with an initial learning rate of 0.0002, which decreases linearly to 0 after 100 iterations.

For training, we manually divided 3661 melanoma images into two types based on the relative size of the lesion. Figure 8 shows some examples of these two types.

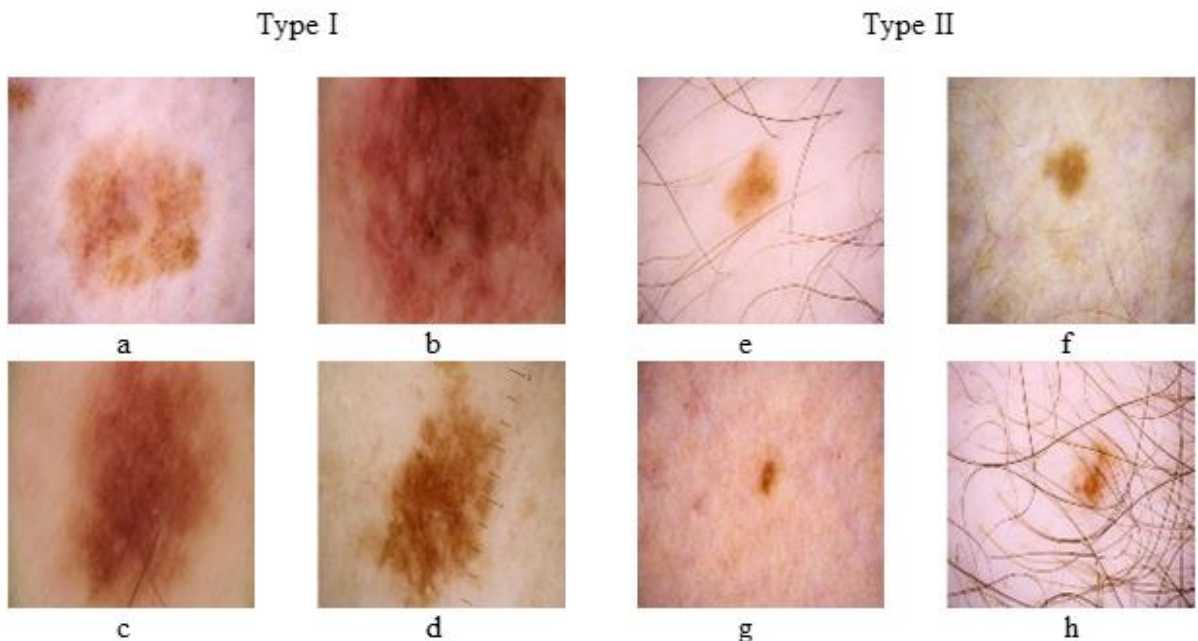

**Fig. 8** Two types of melanoma images based on the lesion coverage. The images a, b, c and d belong to type I as the melanoma covers more than roughly 25 percent of the image. On the contrary, the images e, f, g, and h belong to type II, where the melanoma is relatively small in this type of images.

The classification criterion is the area coverage ratio of melanoma to the image. There are 1597 images in type I, and the area of melanoma is more than 25% of the image. The remaining 2064 melanoma images belong to type II with less than 25% of the images covered by the lesion. These two types of melanoma images are trained with flower images separately. Therefore, we obtain two style transformation models suited to the input image type.

*2.7 Evaluation Methods*

In this work, we evaluate the results from subjective and objective aspects.

The subjective evaluation is mainly based on personal feelings. We select ten melanoma images, convert them with two networks, and obtain ten sets of results to make a questionnaire. Healthy volunteers are asked to choose only one preferred image from each group. In total, 52 randomly selected persons completed the questionnaire.

As objective evaluation criteria, we use the Fréchet Inception Distance (FID)[28] and the structural similarity index measure (SSIM)[29] in this work. The Fréchet Inception Distance (FID) is a metric that evaluates the quality of GAN-generated images by calculating the distance between the real image and the generated image in the feature space through the Inception-v3 network. Lower scores indicate the two groups of images are more similar. The structural similarity index measure (SSIM) evaluates the similarity of images by comparing three features: luminance, contrast, and structure, while incorporating human visual preferences. Higher SSIM values indicate a better perceived quality of the generated images.

## 3  Results

We train the three networks $A$, $B$ and $C$ introduced in the methods section with the available datasets and compare the results as shown in Figure 9. The result $A$ is from the network including the sub-pixel convolution. The result $B$ corresponds to the network structure without sub-pixel convolution. The result $C$ is obtained from the training with the network structure with the additional self-attention and spectral normalization. Figure 9 shows exemplary results for the three GANs.

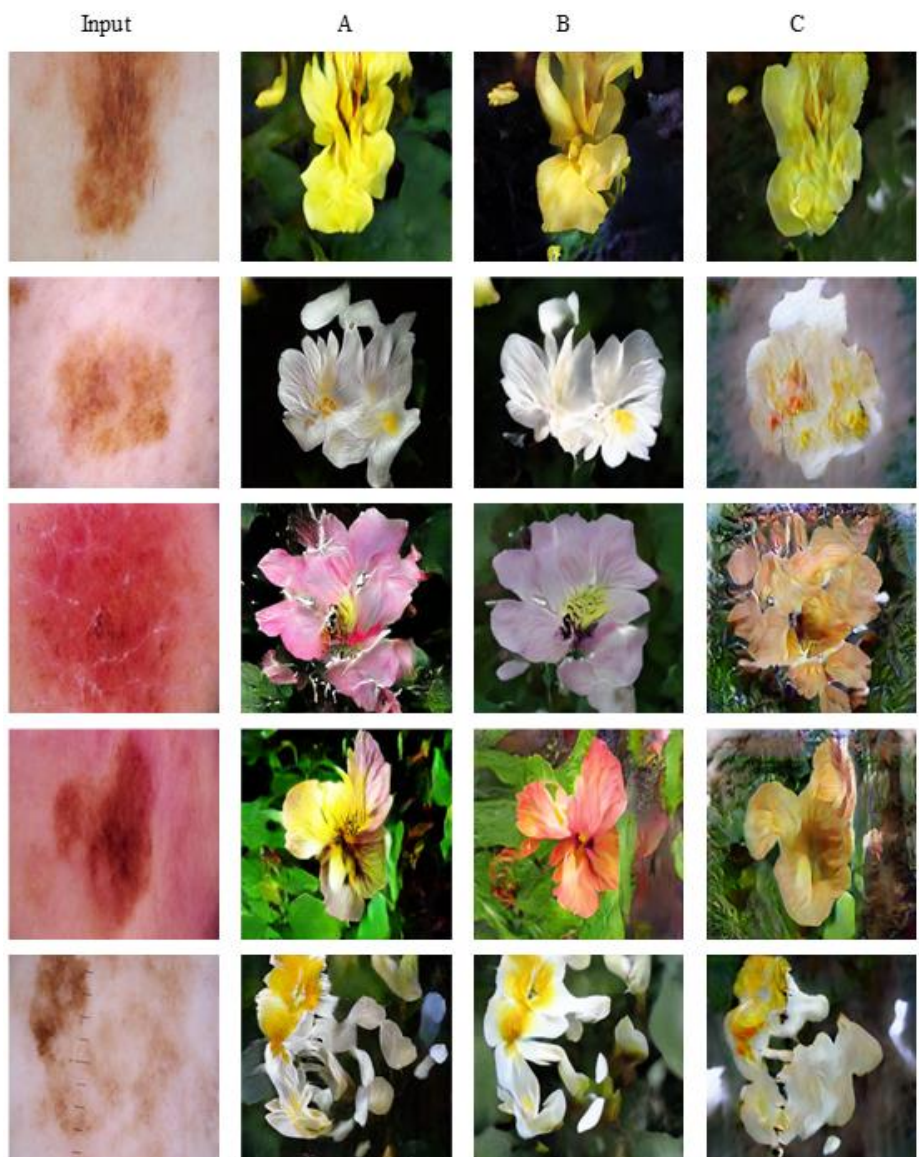

**Fig. 9** Comparison of the three results obtained by training with the three different network structures. The result $A$ is from the network including sub-pixel convolution. The result $B$ corresponds to the network structure introduced without the sub-pixel convolution. The result $C$ is obtained from the training with the network structure containing self-attention and spectral normalization.

From Figure 9 it is obvious that the conversion based on network $C$ delivers less aesthetically pleasing flower images. Therefore, GAN $C$ is not studied further in the remainder of this work.

Since the results of *A* and *B* can not be evaluated by simple comparison, we evaluate the quality of *A* and *B* with our evaluation methods introduced in the methods section.

There are 52 valid responses of the questionnaire from randomly selected healthy volunteers. Figure 10 shows the exemplary results used in the questionnaire.

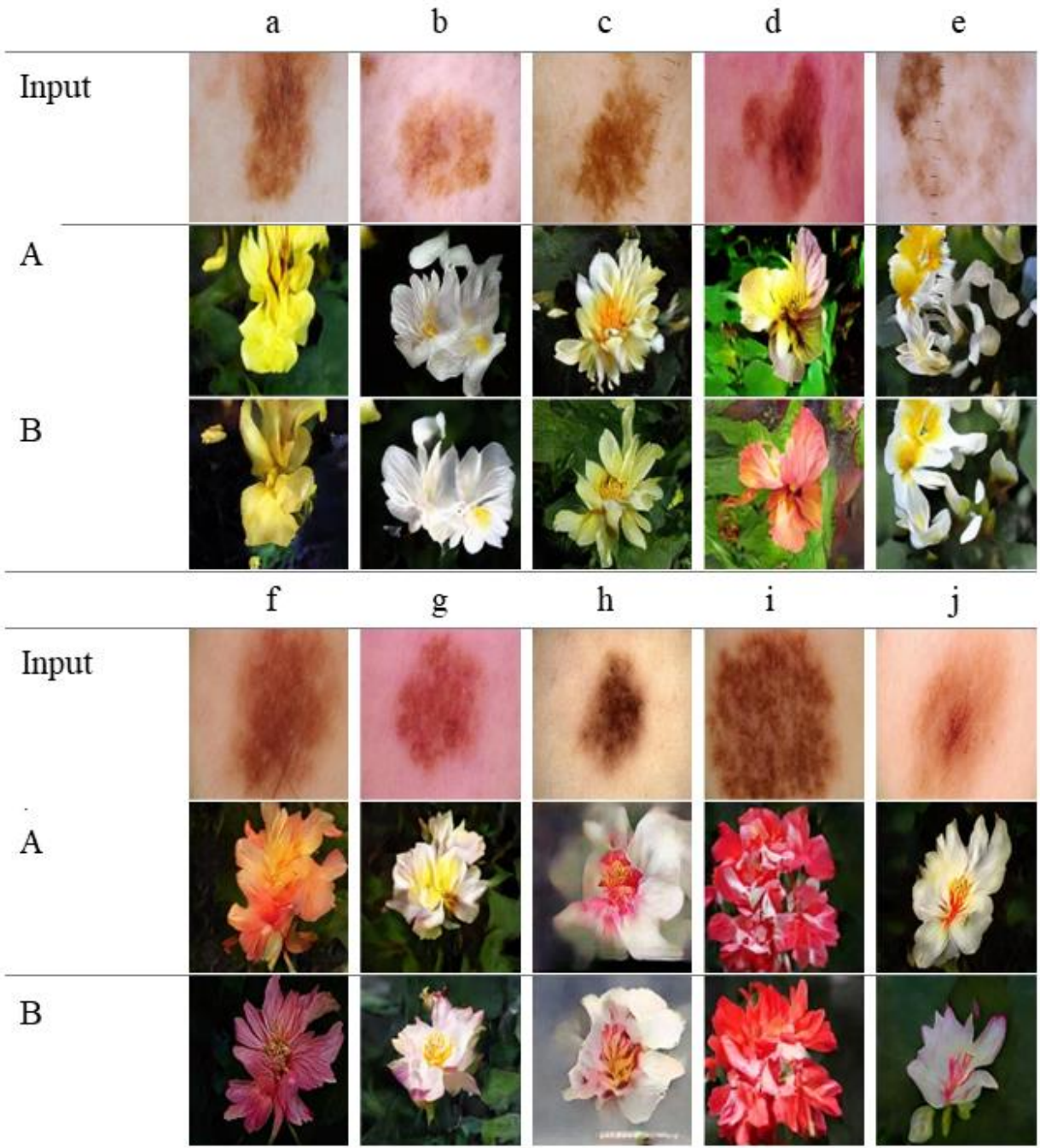

**Fig. 10** Test image result used in the questionnaire. For each image (a-j) 52 healthy volunteers were asked to choose the preferred conversion (A or B).

Table 3 displays the results from the questionnaire and the SSIM values for each image depending on the network type.

**Table 3** Comparison of questionnaire results and SSIM values for each image.

| Example | a | b | c | d | e | f | g | h | i | j |
|---|---|---|---|---|---|---|---|---|---|---|
| **Likes (A)** | 8 | 27 | 22 | 9 | 22 | 35 | 15 | 21 | 22 | 25 |
| **Likes (B)** | 44 | 25 | 30 | 43 | 30 | 17 | 37 | 31 | 30 | 27 |
| **SSIM (A)** | 0.3146 | 0.1836 | 0.2342 | 0.1940 | 0.2488 | 0.3031 | 0.2193 | 0.4594 | 0.3183 | 0.1758 |
| **SSIM (B)** | 0.2141 | 0.2470 | 0.2313 | 0.3924 | 0.3781 | 0.1613 | 0.3098 | 0.5824 | 0.3561 | 0.2773 |

The favorable rating for each result equals to the number of likes divided by the total number of responses. The final SSIM value for each result is the average of the ten SSIM values. Table 4 shows the evaluation values for each group. From the comparison, result *B* is more popular and has higher quality. Furthermore, we calculate the FID scores between the converted two results (*A* and *B*) and the flower images which are also presented in Table 4.

**Table 4** Favorable rating, FID and SSIM value for each result.

| Network | Favorable Rating | FID | SSIM |
|---|---|---|---|
| **A** | 39.62% | 269.17 | 0.26511 |
| **B** | 60.38% | 240.11 | 0.31498 |

As presented in Table 4, 60% of the participants in the questionnaire prefer the conversions from network *B*. Additionally, the lower FID score for network structure *B* indicates that the results from *B* are more similar to the original real flower domain compared to the network structure *A*. Furthermore, the higher SSIM value of network structure *B* shows that *B* provides a better perceived quality of the generated images.

*3.1 User Interface*

We have designed a graphical user interface (GUI) to enable convenient operation of the melanoma image conversion for the relevant personnel.

The GUI of this software is programmed using the PyQt toolkit. A screenshot of the interface is shown in Figure 11.

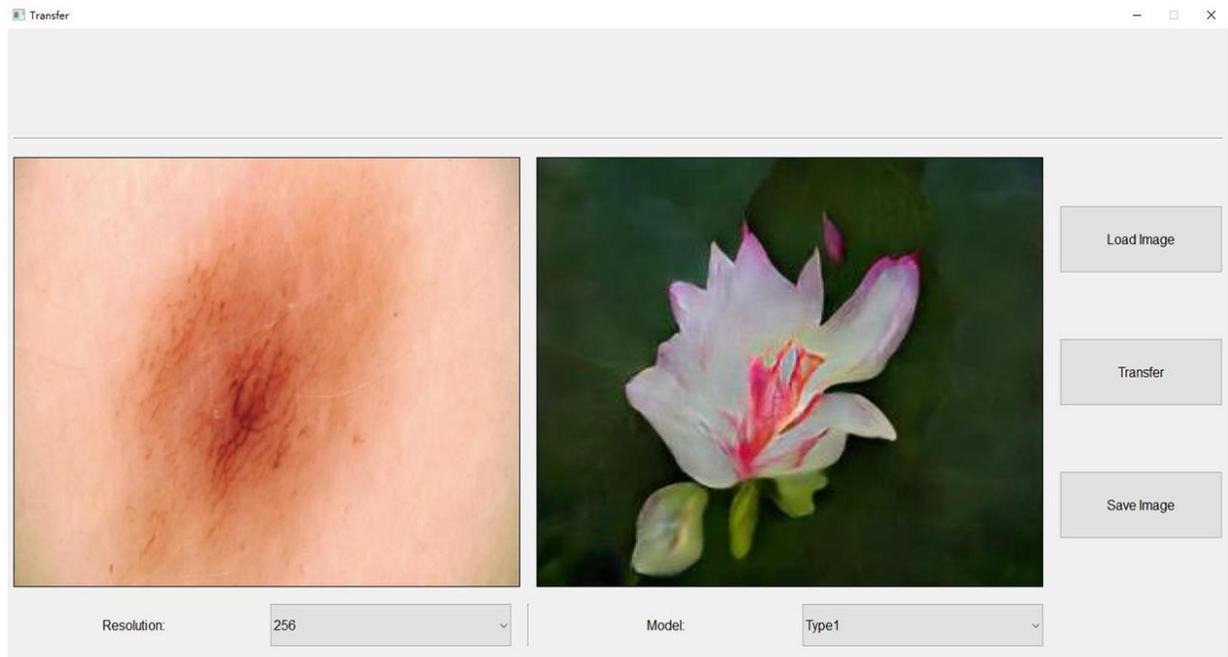

**Fig. 10** Screenshot of the GUI. The functionalities are loading, transfer and saving. Additionally, the operator can choose between the two types of models for conversion proposed in this work and select a resolution.

The software system implements five functions: image loading, image saving, image conversion, model choice, and resolution selection. Users can select the dermoscopic melanoma image they want to convert by clicking *load image* and then choose the resolution and the model. The software has two resolution options, 256 and 512, and it will resize the converted image according to the aspect ratio of the original image. For example, if the input resolution is 6000*4000, the output will be 384*256 or 768*512. The *transfer* function starts the transformation from melanoma to flower, an additional function saves the converted image.

## 4  Conclusion

In this work, we implemented the transformation of melanoma images to flower images utilizing GANs. We trained our datasets utilizing three different network structures based on CycleGAN. The proposed networks can complete the image conversion within seconds. Furthermore, we evaluated the quality of the generated images by subjective evaluation in form of a questionnaire and objective criteria i.e., the parameters SSIM and FID. Based on these results, we determined the final model. Furthermore, we provided a user interface for easy accessibility and usability. In conclusion, we have completed the technical infrastructure that enables, to our knowledge, the first implementation of GANs focusing on patient mental health in melanoma disease management.

Limited by hardware and processing speed, we use images with 256*256 resolution for the network training. In future work, training with better hardware could improve the resolution of the results. It is likely that the resulting networks would deliver more photorealistic images. In addition, as this work focuses on the transformation of melanoma into flowers only, other styles such as clouds or trees can be implemented according to the availability of different datasets. However, natural themes are an intuitive choice as they are known to relieve stress, but also other domains should be addressed especially for young patients. Furthermore, the number of dermoscopic melanoma images as well as flower images in the training datasets used in this work could be increased to improve the image conversion further. It might be conceivable to include this approach of personalized art therapy directly into whole body imaging given recent advancements in complete body dermoscopy for the early diagnosis of melanoma.

*Disclosures*

The authors declare no conflict of interests.

*Acknowledgments*

This work has received funding from the European Union's Horizon 2020 research and innovation programme **iToBoS** under grant agreement No 965221. We also acknowledge funding by the Deutsche Forschungsgemeinschaft (DFG, German Research Foundation) under Germany's Excellence Strategy within the Cluster of Excellence **PhoenixD** (EXC 2122, Project ID 390833453).

*Code, Data, and Materials Availability*

Data is available upon reasonable request.

**Author contributions**

Conceptualization, L.J. and B.R.; methodology, L.J.; software, L.J. and N.W.; validation, L.J. and B.R.; formal analysis, L.J.; investigation, L.J.; resources, B.R.; data curation, L.J. and N.W.; writing—original draft preparation, L.J. and N.W.; writing—review and editing, B.R.; visualization, L.J. and N.W.; supervision, B.R.; project administration, B.R.; funding acquisition, B.R. All authors have read and agreed to the published version of the manuscript.

**Lennart Jütte** is a PhD student at Hannover Centre for Optical Technologies. He received his BS and MS degrees in Nanotechnology from the Leibniz University Hannover in 2019 and 2021, respectively, where he is currently pursuing the PhD degree in applied optics. His current research interests include dermoscopy, biophotonics and polarimetry.

**Bernhard Roth** received the Ph.D. degree in atomic and particle physics from the University of Bielefeld, Bielefeld, Germany, in 2001, and the State Doctorate (Habilitation) degree in experimental quantum optics from the University of Duesseldorf, Duesseldorf, Germany, in 2007. From 2002 to 2007, he was a Research Group Leader with the University of Duesseldorf. Since 2012, he has been the Scientific and Managing Director with the Hanover Centre for Optical Technologies and a Professor of physics with Leibniz University Hanover, Hanover, Germany, since 2014. His scientific activities include applied and fundamental research in laser development and spectroscopy, polymer optical sensing, micro- and nanooptics fabrication, and optical technology for illumination, information technology, and the life sciences.

Biographies and photographs for the other authors are not available.

**Caption List**

**Fig. 1** Cycle consistent GAN model. It contains two generators, $G$ and $F$, and two discriminators, $D_X$ and $D_Y$. Melanoma image as the source domain $X$, flower image as the target domain $Y$.

**Fig. 2** Structural diagram of the generator. The generator network consists of a convolution block, a downsampling block, nine residual blocks, an upsampling block and an output layer.

**Fig. 3** Structural diagram of the discriminator. The discriminator network consists of four convolutional blocks (CONV) that perform convolutional operations to extract image features, a fully connected layer (FCN) to combine the features extracted from the previous convolutional layers, and a global average pooing layer (GAP) to reduce the number of the parameters.

**Fig. 4** Structural diagram of the generator with sub-pixel convolution. Extracting features with a convolutional block and downsampling, residual blocks help to avoid the neural network degradation. We employ sub-pixel convolution to complete the upsampling.

**Fig. 5** Structure of the self-attention mechanism. The three vectors are query ($Q$), key ($K$) and value ($V$). $Q$ is the query vector related to the encoded information, while $K$ is the output vector of the encoder. $V$ is the vector related to the input. With $Q$ and $K$ the attention can be calculated to perform the Softmax and then multiply it with $V$. The input here is the dermoscopic image.

**Fig. 6** Structural diagram of the generator with self-attention. There is one self-attention layer after the downsampling and the residual blocks, respectively.

**Fig. 7** Structural diagram of discriminator with self-attention and spectral normalization. We replaced the instance normalization with spectral normalization and add self-attention after the last two convolution blocks compare to the discriminator introduced in chapter 2.3.

**Fig. 8** Two types of melanoma images based on the lesion coverage. The images a, b, c and d belong to type I as the melanoma covers more than roughly 25 percent of the image. On the contrary, the images e, f, g, and h belong to type II, where the melanoma is relatively small in this type of images.

**Fig. 9** Comparison of the three results obtained by training with the three different network structures. The result *A* is from the network including sub-pixel convolution. The result *B* corresponds to the network structure introduced without the sub-pixel convolution. The result *C* is obtained from the training with the network structure containing self-attention and spectral normalization.

**Fig. 10** Test image result used in the questionnaire. For each image (a-j) 52 healthy volunteers were asked to choose the preferred conversion (A or B).

**Fig. 11** Screenshot of the GUI. The functionalities are loading, transfer and saving. Additionally, the operator can choose between the two types of models for conversion proposed in this work and select a resolution.

**Table 1** The parameters and the size of the output for each convolutional layer of the generator. *N* is the number of channels in the convolutional layer, *K* is the size of the convolution kernel, *S* is the stride, and *P* is the padding size. The input size in this work is 3*256*256(*C* Channel * *H* Height * *W* Width).

**Table 2** The parameters and the size of the output for each convolutional layer of the discriminator, the input size is 3*256*256(Channel*Height*Width). *N* is the number of channels in the convolutional layer, *K* is the size of the convolution kernel, *S* is the stride, and *P* is the padding size. The output size is given as per *C* Channel * *H* Height * *W* Width.

**Table 3** Comparison of questionnaire results and SSIM values for each image.

**Table 4** Favorable rating, FID and SSIM value for each result.

**<u>Appendix</u>**

*A.1    Training Environment*

Table A1 shows the model training environment and configuration in this work. The CPU is Intel(R) Core(TM) i3-9100F, the operating system is windows 10 64-bit, the GPU is NVIDIA GeForce RTX 2060 12GB. Using Pytorch 1.4.0 as a deep learning framework and cuDNN to accelerate the training process on the GPU. CUDA version is 10.1, cuDNN version is 7.6.4, and the programming language is python 3.6.

**Table A1** Environment and Configuration.

| | |
|---|---|
| CPU | Intel(R) Core(TM) i3-9100F |
| GPU | NVIDIA GeForce RTX 2060 12GB |
| OS | Windows 10 64-bit |
| CUDA | 10.1 |
| cuDNN | 7.6.4 |
| Pytorch | 1.4.0 |
| Python | 3.6 |